\documentstyle[aps,preprint,prb]{revtex}
\begin{document}
\centerline      {\sl\bf STRONG COUPLING THEORY OF IMPURE SUPERCONDUCTORS :}
\centerline{\bf REAL SPACE FORMALISM}
\def\hb{\hfill\break}
\vskip 2pc
\centerline {Yong-Jihn Kim$^{\dagger}$}
\vskip 1pc
\centerline {Department of Physics, Purdue University}
\centerline {West Lafayette, Indiana 47907}
\vskip 2pc
In the electron-phonon model, the influence of nonmagnetic impurities
on the transition temperature of superconductors is revisited.  
Anderson's  pairing condition between time-reversed 
eigenstate pairs is 
derived from the physical constraint 
of the Anomalous Green's function.
After this pairing condition is incorporated into the self-consistency
equation, one may find that  
the phonon-mediated interaction decreases exponentially by localization.

\vskip 1.5in\hfill\break
PACS numbers: 74.20.-z, 74.40.+k, 74.60.Mj
\vskip 1.5pc\hfill\break
${\dagger}$ Present address, Department of Physics, Korea Advanced
Institute of 

Science and Technology, Taejon 305-701, Korea.\hfill\break
\vskip 2pc

\vfill\eject 

Impurity scattering can affect the transition temperature of superconductors
by changing the electron density of states and the phonon-mediated
interaction. 
It was shown$^{1}$ that the impurity effect on the electron density of states
is overestimated in Abrikosov and Gor'kov's (AG) theory$^{2}$ of 
impure superconductors. 
In their comment$^{3}$ on ref. 1,\ Abrikosov and Gor'kov argued that their 
theory  
should be reinterpreted in terms of the Eliashberg 
theory$^{4}$ in order to satisfy Anderson's theorem.$^{5}$ 
Their argument follows from Tsuneto's$^{6}$ application of the AG theory 
to the electron-phonon model. 
His result apparently agreed with Anderson's theorem. 
The overestimation didn't occur because of the absence of the
cutoff problem in the Eliashberg theory.

However, the impurity effect on the phonon-mediated 
interaction is missed in both theories.$^{1,7}$
Note that Anderson's theorem does not explain the exponential 
decrease$^{1, 8}$ of the phonon-mediated interaction by localization
because it is valid only to the first order in the impurity 
concentration.$^{1, 7}$
The purpose of this study is to construct a strong coupling theory 
which can take into account the impurity effect on the phonon-mediated
interaction precisely.  
The change of Coulomb interactions and phonon properties
caused by impurities are neglected.$^{9}$

First, Anderson's pairing condition is found 
from the physical constraint of the Anomalous Green's function.$^{10}$ 
Incorporating 
the pairing condition into the self-consistency equation, one finds 
the exponential decrease of the phonon-mediated interaction caused by the
localized wavefunction. 

We follow the real space formalism of the strong 
coupling theory
by Eilenberger and Ambegaokar.$^{11}$ 
(See also refs. 12-16.) 
The Hamiltonian for the electron-phonon interaction takes the form
\begin{eqnarray}
H_{int} = g\int \psi^{+}({\bf r})\psi ({\bf r})\phi({\bf r})d{\bf r}, 
\end{eqnarray}
where $\Psi({\bf r})$ and $\phi({\bf r})$ are the electron and phonon field 
operators. $g$ is the coupling constant.
The equations of the motion for the thermodynamic Green's functions
$G(\omega_{n}, {\bf r}, {\bf r'})$ and $F^{+}(\omega_{n}, {\bf r}, {\bf r'})$
are given
\begin{eqnarray}
(i\omega_{n} + {1\over 2m}\nabla^{2} + V({\bf r}) + \mu)
G(\omega_{n}, {\bf r}, {\bf r'})  &=& \delta({\bf r}-{\bf r'}) + g^{2}\int 
d{\bf r}_{o}
\Sigma(\omega_{n}, {\bf r},{\bf r}_{o}) G(\omega_{n}, {\bf r}_{o}, {\bf r'})\nonumber \\
&+& g^{2}\int d{\bf r}_{o} \phi(\omega_{n}, {\bf r},{\bf r}_{o})
F^{+}(\omega_{n}, {\bf r}_{o}, {\bf r'}), 
\end{eqnarray}
\begin{eqnarray}
(-i\omega_{n} + {1\over 2m}\nabla^{2} + V({\bf r}) + \mu)
F^{+}(\omega_{n}, {\bf r}, {\bf r'})  &=&  g^{2}\int d{\bf r}_{o}
\Sigma^{+}(\omega_{n}, {\bf r},{\bf r}_{o})
F^{+}(\omega_{n}, {\bf r}_{o}, {\bf r'})
\nonumber\\
&+& g^{2}\int d{\bf r}_{o}
\phi^{+}(\omega_{n}, {\bf r},{\bf r}_{o})
G(\omega_{n}, {\bf r}_{o}, {\bf r'}),
\end{eqnarray}
where 
\begin{eqnarray}
 \Sigma(\omega_{n}, {\bf r},{\bf r'}) = T\sum_{n'} D(\omega_{n}, \omega_{n'},
{\bf r}, {\bf r'})G(\omega_{n'}, {\bf r}, {\bf r'}), 
\end{eqnarray}
\begin{eqnarray}
 \phi(\omega_{n}, {\bf r},{\bf r'}) = T\sum_{n'} D(\omega_{n}, \omega_{n'},
{\bf r}, {\bf r'})F^{+}(\omega_{n'}, {\bf r}, {\bf r'}). 
\end{eqnarray}
$V({\bf r})=\sum_{i}V_{o}\delta({\bf r}-{\bf R}_{i})$ is the scattering potential 
of the impurities and
$\omega_{n} = (2n+1)\pi T$. $D$ is the phonon Green's function.

It is usually assumed that the electron-phonon interaction is 
local,$^{4, 11, 12}$
(i.e.),
\begin{eqnarray}
(\omega_{n},\omega_{n'}, {\bf r},{\bf r'}) = \delta({\bf r}-{\bf r'})
\lambda(\omega_{n},\omega_{n'}),
\end{eqnarray}
\begin{eqnarray}
 \Sigma(\omega_{n}, {\bf r},{\bf r'}) = 
 \delta({\bf r}-{\bf r'}) \Sigma(\omega_{n}, {\bf r}), 
\end{eqnarray}
\begin{eqnarray}
 \phi(\omega_{n}, {\bf r},{\bf r'}) = 
 \delta({\bf r}-{\bf r'}) 
 \phi(\omega_{n}, {\bf r}). 
\end{eqnarray}
The generalization to the non-local interactions is straightforward and
will be published elsewhere.

The normal-state Green's function $G_{N}(\omega_{n},{\bf r},{\bf r'})$
satisfies the equation
\begin{eqnarray}
(i\omega_{n} + {1\over 2m}\nabla^{2} + V({\bf r}) + \mu)
G_{N}(\omega_{n}, {\bf r}, {\bf r'})  = \delta({\bf r}-{\bf r'}) + g^{2}\int 
d{\bf r}_{o} \Sigma(\omega_{n}, {\bf r},{\bf r}_{o})
G_{N}(\omega_{n}, {\bf r}_{o}, {\bf r'}) .
\end{eqnarray}
From Eqs. (3) and (9) the Anomalous Green's 
function $F^{+}(\omega_{n}, {\bf r},{\bf r'})$, near the transition temperature,
 can be rewritten in the form
\begin{eqnarray}
 F^{+}(\omega_{n}, {\bf r},{\bf r'}) = g^{2}\int d{\bf r}_{o}G_{N}(-\omega_{n},
{\bf r}_{o},{\bf r})\phi^{+}(\omega_{n},{\bf r}_{o})G_{N}(\omega_{n},{\bf r}_{o},{\bf r'}).
\end{eqnarray}
Accordingly, we obtain the self-consistency equation for $\phi^{+}$
\begin{eqnarray}
\phi^{+}(\omega_{n}, {\bf r}) &=& T\sum_{n'}\lambda
(\omega_{n},\omega_{n'}) F^{+}(\omega_{n'}, {\bf r},{\bf r})\nonumber\\
&=& g^{2}T\sum_{n'}\lambda(\omega_{n},\omega_{n'})\int d{\bf r}_{o}G_{N}(-\omega_{n'},{\bf r}_{o},
{\bf r})G_{N}(\omega_{n'},{\bf r}_{o},{\bf r})\phi^{+}(\omega_{n'},{\bf r}_{o}).
\end{eqnarray}
In the presence of the impurities, the normal-state Green's function
$G_{N}$ may be given by
\begin{eqnarray}
G_{N}(\omega_{n}, {\bf r},{\bf r'}) = \sum_{m}{\psi_{m}({\bf r})\psi^{*}_{m}
({\bf r'})\over i\omega_{n}Z(\omega_{n})-\epsilon_{m}}, 
\end{eqnarray}
where $Z(\omega_{n})$ is the renormalization factor and $\psi_{m}({\bf r})$ is the 
scattered eigenstate.

The pair potential $\Delta^{*}(\omega_{n},{\bf r})$ is defined by
 $\Delta^{*}(\omega_{n},{\bf r}) = \phi^{+}(\omega_{n},{\bf r}) / Z(\omega_{n})$.
Therefore we find the self-consistency equation for the pair potential
to be$^{11}$
\begin{eqnarray}
\Delta^{*}&(&\omega_{n}, {\bf r})Z(\omega_{n})  = \nonumber\\
 &g^{2}&T\sum_{n'}\lambda(\omega_{n},\omega_{n'})\int d{\bf r}_{o}G_{N}(-\omega_{n'},{\bf r}_{o},
{\bf r})G_{N}(\omega_{n'},{\bf r}_{o},{\bf r})\Delta^{*}(\omega_{n'},{\bf r}_{o})
Z(\omega_{n'}).  
\end{eqnarray}
Let's compare this with the self-consistency equation in the Gor'kov's 
formalism which is given:
\begin{eqnarray}
\Delta^{*}({\bf r}) = VT\sum_{\omega}\int 
G(-\omega,{\bf r}_{o},{\bf r})
G(\omega, {\bf r}_{o},{\bf r})\Delta^{*}({\bf r}_{o})d{\bf r}_{o}.
\end{eqnarray}
Eq. (13) has the additional frequency dependence in the kernel.
Notice that the spatial parts of the kernels in both equations are the same.

However, it was pointed out$^{10}$ that Eq. (14) allows extra pairings
which violate the physical constraint of the Anomalous Green's function.
If we substitute Eq. (12) (with $Z=1$) into Eq. (14), 
there are extra pairings between $m\uparrow$ and $m'\downarrow$ as well as
Anderson's pairing between $m\uparrow$ and ${\bar m}\downarrow$.
$\bar m$ denotes the time reversed partner of the
scattered state $m$.
$m'\downarrow$ does not include ${\bar m}\downarrow$.
Because of the same spatial parts of the kernel, 
Eq. (13) also gives rise to extra pairings.
In the Bogoliubov-de Gennes equations, the situations are more clear.$^{17}$
The corresponding unitary transformation leads to
the vacuum state where pairing occurs between the states which
are the linear combination of the scattered states.
A similar problem was found in Gor'kov and Galitski's (GG)$^{18}$ solution
for the d-state BCS theory. GG allowed a superposition
of several distinct types of the off-diagonal-long-range-order.$^{19,20,21}$

Now we check the physical constraint of the Anomalous Green's function.
If we average over the impurity positions, the Anomalous Green's function
should  be a function of ${\bf r}-{\bf r'}$ due to the recovery of homogeneity.
Using the scattered states $\psi_{m(\vec k)}(\bf r)$,
\begin{eqnarray}
\psi_{m(\vec k)}({\bf r}) = e^{i\vec k\cdot{\bf r}} + \sum_{\vec q}{V_{o}\over \epsilon_{\vec k} -
\epsilon_{\vec k+\vec q}}[\sum_{i}e^{-i\vec q\cdot{\vec R_{i}}}]e^{i(\vec k+\vec q)\cdot {\bf r}}
+  \cdots ,
\end{eqnarray}
it is easy to show that
\begin{eqnarray}
\overline{\psi_{m(\vec k)\uparrow}({\bf r})\psi_{m'(\vec k')\downarrow}({\bf r'})}^{imp} 
 &=& e^{i(\vec k\cdot{\bf r}+\vec k'\cdot{\bf r '})} + V_{o}^{2}f'({\bf r},{\bf r'}) + \cdots \nonumber\\
 &\not=& f({\bf r}-{\bf r'}),
 \end{eqnarray}
and
\begin{eqnarray}
\overline{\psi_{m(\vec k)\uparrow}({\bf r})\psi_{{\bar m}(-\vec k)\downarrow}({\bf r'})}^{imp} 
 &=& e^{i\vec k\cdot({\bf r}-{\bf r '})} + V_{o}^{2}g'({\bf r}-{\bf r'}) + \cdots 
 \nonumber\\
 &\equiv& g({\bf r}-{\bf r'}).
 \end{eqnarray}
$\bar{\ }\bar{\ }^{imp}$ means an average over impurity positions
$\vec R_{i}$.
$f, f', g,$ and $g'$ can be readily calculated.
Note that even the zeroth order term of extra pairings between $m\uparrow$ and  
$m'(\not= {\bar m})\downarrow$ is not a function of ${\bf r}-{\bf r'}$.
To calculate the impurity average of the Anomalous Green's function,
we need to consider the following average:
\begin{eqnarray}
\overline{F^{+}(\omega_{n},{\bf r},{\bf r'})}^{imp}&\sim& 
\overline{\psi_{m\uparrow}^{*}({\bf r})\psi_{m'\downarrow}^{*}({\bf r'})\psi_{m\uparrow}({\bf r}_{o})
\psi_{m'\downarrow}({\bf r}_{o})\phi^{+}({\bf r}_{o})}^{imp}\nonumber\\
&=& \overline{\psi_{m\uparrow}^{*}({\bf r})\psi_{m'\downarrow}^{*}({\bf r'})}^{imp}\ \
\overline{\psi_{m\uparrow}({\bf r}_{o})
\psi_{m'\downarrow}({\bf r}_{o})\phi^{+}({\bf r}_{o})}^{imp}\nonumber\\
&+& {\rm correction \ terms}.
\end{eqnarray}
From Eqs. (16)-(18), it is straight forward to show that
 \begin{eqnarray}
\overline{F^{+}(\omega_{n},{\bf r},{\bf r'})}^{imp} \not= 
\overline{F^{+}(\omega_{n},{\bf r}-{\bf r'})}^{imp},
 \end{eqnarray}
 \begin{eqnarray}
\overline{\Delta^{*}(\omega_{n},{\bf r})}^{imp} \not= \Delta^{*}(\omega_{n}),
 \end{eqnarray}
in the presence of the extra pairing terms.
Only Anderson's pairing between $m\uparrow$ and ${\bar m}\downarrow$ is compatible
with the homogeneity condition of the Anomalous Green's function
after averaging out the impurity positions.

Consequently, the self-consistency equation needs a proper pairing
constraint derived from the Anomalous Green's function.
The resulting equation is
\begin{eqnarray}
\Delta^{*}&(&\omega_{n}, {\bf r})Z(\omega_{n})  = \nonumber\\
 &g^{2}&T\sum_{n'}\lambda(\omega_{n},\omega_{n'})\int d{\bf r}_{o}
\{G_{N}(-\omega_{n'},{\bf r}_{o}, {\bf r})
G_{N}(\omega_{n'},{\bf r}_{o},{\bf r})\}_{p.p.}
\Delta^{*}(\omega_{n'},{\bf r}_{o})Z(\omega_{n'}),  
\end{eqnarray}
where
$p.p.$ denotes the {\sl proper pairing constraint}. 
For ordinary impurities, Anderson's pairing condition between the
time-reversed  partners is obtained. 
The gap parameter $\Delta^{*}(\omega_{n}, m)$ is given by$^{22}$
\begin{eqnarray}
\Delta^{*}(\omega_{n}, m) = \int \psi_{m}({\bf r})\psi^{*}_{m}({\bf r})
\Delta^{*}(\omega_{n},{\bf r})d{\bf r}.
\end{eqnarray}
Finally, we find a gap equation
\begin{eqnarray}
\Delta^{*}(\omega_{n}, m) = 
\sum_{n'}\lambda(\omega_{n},\omega_{n'})
 \sum_{m'}V_{mm'}{\Delta^{*}(\omega_{n'},m')\over [-i\omega_{n'}Z(\omega_{n'})
-\epsilon_{m'}][i\omega_{n'}Z(\omega_{n'})-\epsilon_{m'}]},
\end{eqnarray}
where
\begin{eqnarray}
V_{mm'} = g^{2}\int |\psi_{m}({\bf r})|^{2}
 |\psi_{m'}({\bf r})|^{2}d{\bf r}.
\end{eqnarray}
As a rule, the same result may be obtained by the k-space formalism.$^{4}$

On the other hand, Tsuneto$^{6}$ obtained the gap equation
\begin{eqnarray}
\Sigma_{2}(\omega) = {i\over (2\pi)^{3}p_{o}}\int dq\int d\epsilon \int 
d\omega ' {qD(q,\omega-\omega')\eta(\omega')\Sigma_{2}(\omega')\over \epsilon^{2} - 
\eta^{2}(\omega')\omega'^{2}},
\end{eqnarray}
where $\eta=1 +{1\over 2\tau|\omega|}$, $\omega=i\omega_{n}$,  and $\tau$ is the collision time.
Comparing Eqs. (23) and (25), we find that Tsuneto's result misses
the most important factor $V_{mm'}$, which gives the change of the
phonon-mediated interaction due to the impurities. 
Anderson's theorem also assumes $V_{mm'} = g^{2}\times 1$.$^{1}$
It is clear that this factor is exponentially small for the localized
states, since the overlap of two localized states
 should in fact be exponentially small.$^{1}$ 
Let's consider the states $m$ and $m'$ localized at ${\bf r}_{m}$
and ${\bf r}_{m'}$, respectively. 
The localization lengths are taken to be $\xi$ and $\xi'(\sim \xi)$.
The  overlap of these wavefunctions are roughly proportional to
$|\psi_{m}({\bf r}_{m}+{{\bf r}_{m}-{\bf r}_{m'}\over 2 })|^{2}$ multipied by
volume of the overlap region.
It may be estimated that 
\begin{eqnarray}
V_{mm'} \sim e^{-{|{\bf r}_{m}-{\bf r}_{m'}|\over 4\xi}}{\xi^{d}}, 
\end{eqnarray}
where $d$ is the dimensionality of the system.
Weak localization correction to the phonon-mediated interaction
will be published elsewhere.

In conclusion, we have constructed a strong coupling
theory of impure superconductors. 
Anderson's  pairing condition, obtained from the physical constraint
of the Anomalous Green's function, is incorporated into the
self-consistency equation.
One finds that the phonon-mediated interaction
decreases exponentially by localization.

I am grateful to Professor A. W. Overhauser for valuable discussions.
This investigation was motivated by the response of Professor K. J. Chang 
at KAIST to my seminar on ref. 10 on Nov. 19, 1994.
This work was supported by the National Science Foundation, Materials Theory
Program.
\vfill\eject
\centerline      {\bf REFERENCES} 
\vskip 1pt\hfill\break
1. Yong-Jihn Kim and A. W. Overhauser, Phys. Rev. B{\bf 47}, 8025 (1993).\hb
2. A. A. Abrikosov and L. P. Gor'kov, Sov. Phys. JETP {\bf 12}, 1243 (1961).\hb
3. A. A. Abrikosov and L. P. Gor'kov, Phys. Rev. B{\bf 49}, 12337 (1994).\hb
4. G. M. Eliashberg, Sov. Phys. JETP {\bf 11}, 696 (1960).\hb
5. P. W. Anderson, J. Phys. Chem. Solids {\bf 11}, 26 (1959).\hb
6. T. Tsuneto, Prog. Theo. Phys. {\bf 28}, 857 (1962).\hb
7. Yong-Jihn Kim and A. W. Overhauser, Phys. Rev. B{\bf 49}, 15779 (1994).\hb
8. G. D. Mahan, {\sl Many-Particle Physics}, (Plenum, New York, 1981), p. 38.\hb
9. D. Belitz, Phys. Rev. B {\bf 35}, 1636 (1987). This paper contains a review of

 prior studies.\hb
10. Yong-Jihn Kim, unpublished.\hb
11. G. Eilenberger and V. Ambegaokar, Phys. Rev. {\bf 158}, 332 (1967).\hb
12. N. F. Masharov, Sov. Phys. Solid State, {\bf 16}, 1524 (1975).\hb 
13. A. A. Abrikosov, L. P. Gor'kov, and I. E. Dzyaloshinski, {\sl Methods of
Quantum Field 

Theory in Statistical Physics} (Prentice-Hall, Englewood, NJ, 1963), 
Sec. 35.\hb
14. N. Menyhard, Nuovo Cimento, {\bf 44}, 213 (1966).\hb
15. E. D. Yorke and A. Bardasis, Phys. Rev. {\bf 159}, 344 (1967).\hb
16. N. R. Werthamer and W. L. McMillan, Phys. Rev. {\bf 158}, 415 (1967).\hb
17. Yong-Jihn Kim, unpublished.\hb
18. L. P. Gor'kov and V. M. Galitskii, Sov. Phys. JETP {\bf 13}, 792 (1961).\hb
19. D. Hone, Phys. Rev. Lett. {\bf 8}, 370 (1962).\hb
20. R. Balian, L. H. Nosanow, and N. R. Werthamer, Phys. Rev. {\bf 8}, 372 (1962).\hb
21. P. W. Anderson, Bull. Am. Phys. Soc. {\bf 7}, 465 (1962); Rev. Mod. Phys.
{\bf 38}, 

298 (1966).\hb
22. M. Ma and P. A. Lee, Phys. Rev. B{\bf 32}, 5658 (1985).
		   
\end{document}